\documentclass[10pt, conference, a4paper, oneside, twocolumn]{IEEEtran}
\IEEEoverridecommandlockouts
\usepackage{lipsum}
\usepackage{graphicx}
\usepackage{amsmath}
\usepackage{amssymb}
\usepackage{mathrsfs}
\usepackage{stfloats}
\usepackage{dsfont}
\usepackage{setspace}
\usepackage{multirow}
\begin{document}
\title{On Spectrum and Infrastructure Sharing in Multi-Operator Cellular Networks}
\author{\authorblockN{Shanshan Wang$^{(1)}$, Konstantinos Samdanis$^{(2)}$, Xavier Costa Perez$^{(2)}$, and Marco Di Renzo$^{(1)}$} \medskip
\authorblockA{$^{(1)}$ Universit\'{e} Paris Saclay -- Laboratoire des Signaux et Syst\`emes \\
CNRS -- CentraleSup\'{e}lec -- Universit\'{e} Paris Sud \\
3 rue Joliot-Curie, 91192 Gif-sur-Yvette (Paris), France \\
e-mail: \{shanshan.wang, marco.direnzo\}@l2s.centralesupelec.fr\medskip \\
$^{(2)}$ NEC Laboratories Europe\\
Kurfursten-Anlage 36, 69115 Heidelberg, Germany\\
e-mail: \{konstantinos.samdanis, xavier.costa\}@neclab.eu}
}
\maketitle
\begin{abstract}
In this paper, we introduce a mathematical framework for analyzing and optimizing multi-operator cellular networks that are allowed to share spectrum licenses and infrastructure elements. The proposed approach exploits stochastic geometry for modeling the locations of cellular base stations and for computing the aggregate average rate. The trade-offs that emerge from sharing spectrum frequencies and cellular base stations are quantified and discussed.
\end{abstract}
\section{Introduction} \label{Introduction}
The number of mobile subscribers accessing cellular networks is experiencing a rapid growth. It is expected that 5.13 billion people worldwide will use a mobile phone in 2017, apart from tablets, laptop and other mobile devices. It is predicted that nearly three-fourths of the world mobile data traffic will be video by 2019, which puts a greater pressure on data-intensive applications \cite{andrews2014will}.

In this context, spectrum sharing and infrastructure sharing are considered to be good solutions to utilize the existing resources more efficiently. They allow, in fact, the sharing of resources among telecommunication operators and provide integrated mobile data services for mobile subscribers \cite{DaSilva}. Infrastructure sharing indicates the possibility of accessing, \textit{e.g.}, physical Base Stations (BSs) that belong to different operators. The users can be connected to the BSs of different operators, while the spectrum resources that they use can be different. Considering conventional cellular networks as an example, the frequency spectrum is divided into frequency bands and different operators transmit over different frequency bands. In spectrum sharing, different operators have the possibility of using the frequency bands of other operators.

In \cite{jorswieck2014spectrum}, the gain, in terms of network efficiency, of spectrum sharing between mobile operators is explored. Spectrum sharing is classified into orthogonal spectrum sharing and non-orthogonal spectrum sharing. The difference lies in the fact that the frequency bands can or cannot be allocated simultaneously to two operators in the former and in the latter case, respectively. In \cite{janssen2014expiration}, the authors quantify the benefits that spectrum sharing brings to cellular systems when experiencing peak loads, and estimate the evolution of these benefits as the traffic load and the
spectrum availability change. In \cite{peha2009sharing}, spectrum sharing is applied in cognitive radio systems, where different spectrum and infrastructure sharing modes for primary and secondary users are considered.

In this paper, the potential of spectrum and infrastructure sharing is investigated from the system-level standpoint. The BSs of different operators are modeled as points of a Poisson Point Process (PPP) and tools from stochastic geometry are used to compute relevant performance metrics \cite{MDR_TCOMrate}, \cite{lu2015stochastic}. Further information on stochastic geometry modeling of cellular networks is available in \cite{MDR_COMML2014}-\cite{WeiRelays_TCOM}. Numerical results are shown in order to quantify the advantages of spectrum and infrastructure sharing for typical cellular network deployments. The proposed approach is general and can be applied to conventional cellular networks and to millimeter-wave cellular networks \cite{MDR__mmWave}.

This paper is organized as follows. In Section \ref{SystemModel}, the system model is introduced. In Section \ref{ASE_RATE_Analysis}, a mathematical framework to compute the aggregate average rate of cellular networks with spectrum and infrastructure sharing is introduced. In Section \ref{Results}, numerical simulations are shown. Finally, Section \ref{Conclusion} concludes this paper.

\textit{Notation}: ${\cal H}(x)$ is the Heaviside step function and $\overline {\cal H} = 1 - {\cal H}(x)$. $\mathbb{E} \{ \cdot \} $ is the expectation operator. ${}_2{F_1}\left( \cdot, \cdot, \cdot, \cdot \right)$ is the Gauss hypergeometric function. ${\rm{MGF}}_X(z)$ is the Moment Generating Function (MGF) of the random variable X, \textit{i.e.}, ${\rm{MGF}}_X(z) = \mathbb{E}{_X}\left\{ {\exp ( - zX)} \right\}$.
\begin{figure*}[!t]
\setcounter{equation}{5}
\begin{equation}
\label{eq_inten1}
{\Lambda _S}([0,x)) = \pi \lambda q_S^{[0,D)}{\left(\frac{x}{{{k}}}\right)^{\frac{2}{{{\alpha _S}}}}}\overline {\cal H} (x - {k}{D^{{\alpha _S}}}) + \pi \lambda \left( {{D^2}(q_S^{[0,D)} - q_S^{[D,\infty )}) + q_S^{[D,\infty )}{{\left(\frac{x}{{{k}}}\right)}^{\frac{2}{{{\alpha _S}}}}}} \right){\cal H}(x - {k}{D^{{\alpha _S}}})
\end{equation}
\normalsize \hrulefill \vspace*{-0.0cm}
\end{figure*}
\begin{figure*}[!t]
\setcounter{equation}{6}
\begin{equation}
\label{eq_inten2}
\widehat \Lambda _S([0,x)) = \frac{{2\pi \lambda }}{{x{\alpha _S}}}{\left( {\frac{x}{{{k}}}} \right)^{\frac{2}{{{\alpha _S}}}}}\left( {q_S^{[0,D)}\overline {\cal H} (x - {k}{D^{{\alpha _S}}}) + q_S^{[D,\infty )}{\cal H}(x - {k}{D^{{\alpha _S}}})} \right)
\end{equation}
\normalsize \hrulefill \vspace*{-0.0cm}
\end{figure*}
\section{System Model} \label{SystemModel}
A downlink cellular network is considered. The BSs of each operator are deployed according to a homogeneous PPP, denoted by ${\Psi}$, of density $\lambda$. It is assumed that full traffic load conditions hold, \textit{i.e.}, all the available BSs are activated. The analysis is conducted for the typical Mobile Terminal (MT) of each operator \cite{MDR_TCOMrate}. The intended link is identified by the superscript $(0)$.
\subsection{Path Loss Model} \label{pathlossmodel}
According to \cite{lu2015stochastic} and \cite{WLu__mmWave}, we consider a path-loss model that accounts for Line-Of-Sight (LOS) and Non-Line-Of-Sight (NLOS) links. This allows us to develop a mathematical framework that can be used for both conventional and millimeter-wave cellular networks \cite{MDR__mmWave}.

For ease of description, a two-ball link state model is considered \cite{lu2015stochastic}. More precisely, the probability of a link of length $r$ to be in LOS/NLOS can be formulated as follows:
\setcounter{equation}{0}
\begin{equation}
\label{eq_oneball}
{p_S}(r) = \left\{ {\begin{array}{*{20}{c}}
{q_S^{[0,D)}}&{r \in [0,D)}\\
{q_S^{[D,\infty )}}&{r \in [D,\infty )}
\end{array}} \right.
\end{equation}
where $q_{\rm{LOS}}^{\left[ {0,D} \right)}$ and $q_{\rm{NLOS}}^{\left[ {0,D} \right)}$ are the probabilities that the link is in LOS and in  NLOS inside the ball of radius $D$, and $q_{\rm{LOS}}^{\left[ {D,\infty } \right)}$ and $q_{\rm{NLOS}}^{\left[ {D,\infty } \right)}$ are the probabilities that the link is in LOS and in  NLOS outside the ball of radius $D$. By definition, $q_{\rm{LOS}}^{\left[ {0,D} \right)}+q_{\rm{NLOS}}^{\left[ {0,D} \right)}=1$ and $q_{\rm{LOS}}^{\left[ {D,\infty } \right)}+q_{\rm{NLOS}}^{\left[ {D,\infty } \right)}=1$.

The path-loss function is defined as follows:
\setcounter{equation}{1}
\begin{equation}
\label{eq_pathloss}
{l_S}(r) = {k}{r^{{\alpha _S}}}
\end{equation}
where $k$ and ${\alpha _S}$ for $S \in \{ {\rm{LOS}}, {\rm{NLOS}}\}$ are the path-loss constant and the path-loss exponent, respectively.

Based on this model, the BSs of each operator can be split in two sets denoted by ${\Psi _{\rm{LOS}}}$ and ${\Psi _{\rm{NLOS}}}$, where $\Psi = {\Psi _{\rm{LOS}}} \cup{\Psi _{\rm{NLOS}}}$. Based on the thinning theorem of PPPs, ${\Psi _{\rm{LOS}}}$ and ${\Psi _{\rm{NLOS}}}$ are two independent and non-homogeneous PPPs of density ${\lambda _{\rm{LOS}}(r)} = \lambda {p_{\rm{LOS}}(r)}$ and ${\lambda _{\rm{NLOS}}(r)} =\lambda {p_{\rm{NLOS}}(r)}$.

The channels are assumed to be distributed according to a Rayleigh fading model with unit power.
\subsection{Base Station Association}
The BS serving the typical MT is chosen according to the smallest path-loss criterion. Let $L_{\rm{LOS}}^{(0)}$ and $L_{\rm{NLOS}}^{(0)}$ denote the smallest path-loss of the PPPs of LOS and NLOS links defined, for $S \in \{ {\rm{LOS}}, {\rm{NLOS}}\}$, as follows:
\setcounter{equation}{2}
\begin{equation}
\label{eq_smallLNL}
L_{S}^{(0)} = {\min _{n \in {\Psi _{S}}}}\{ {l_{S}}({r_{n}})\}
\end{equation}
where $r_{n}$ is the distance from a generic BS to the typical MT. Then, the smallest path-loss can be formulated as ${L^{(0)}} = \min \{ L_{\rm{LOS}}^{(0)},L_{\rm{NLOS}}^{(0)}\}$.

By using the same approach as in \cite{MDR__mmWave}, the Cumulative Distribution Function (CDF) and the Probability Density Function (PDF) of ${L^{(0)}}$ can be formulated as follows:
\setcounter{equation}{3}
\begin{equation}
\label{eq_CCDFPDF}
\begin{aligned}
{{\rm{CDF}}_{{L^{(0)}}}}(x) &= 1 - \exp ( - \Lambda ([0,x)))\\
{{\rm{PDF}}_{{L^{(0)}}}}(x) &= {\widehat \Lambda}([0,x))\exp ( - \Lambda ([0,x)))
\end{aligned}
\end{equation}
where the following definitions hold:
\setcounter{equation}{4}
\begin{equation}
\label{eq_intensity}
\begin{aligned}
\Lambda ([0,x)) &= {\Lambda _{\rm{LOS}}}([0,x)) + {\Lambda _{\rm{NLOS}}}([0,x))\\
{\widehat \Lambda}([0,x)) &= \widehat \Lambda _{\rm{LOS}}([0,x)) + \widehat \Lambda _{\rm{NLOS}}([0,x))
\end{aligned}
\end{equation}
where ${\Lambda _S}(\cdot,\cdot)$ and $\widehat \Lambda _S(\cdot,\cdot)$ for $S \in \{ {\rm{LOS}}, {\rm{NLOS}}\}$ are provided in \eqref{eq_inten1} and \eqref{eq_inten2} shown at the top of this page.
\begin{figure*}[!t]
\setcounter{equation}{10}
\begin{equation}
\label{eq_MGF}
\begin{split}
& {\rm{\overline {MGF}}}_{I,i,S}(z;x) \\
&= \exp \left( {\pi \lambda_i k^{ - \frac{2}{{{\alpha _S}}}}{x^{\frac{2}{{{\alpha _S}}}}}\left( {1 - {}_2{F_1}\left( { - \frac{2}{{{\alpha _S}}},1,1 - \frac{2}{{{\alpha _S}}};\frac{{ - z}}{{x}}} \right)} \right)\left( {q_S^{[0,D)}\overline {\cal H}\left( {x - {k}{D^{{\alpha _S}}}} \right) + q_S^{[D,\infty )}{\cal H}\left( {x - {k}{D^{{\alpha _S}}}} \right)} \right)} \right)\\
 & \times \exp \left( {\pi \lambda_i \left( {q_S^{[D,\infty )} - q_S^{[0,D)}} \right){D^2}\left( {1 - {}_2{F_1}\left( { - \frac{2}{{{\alpha _S}}},1,1 - \frac{2}{{{\alpha _S}}};\frac{{ - z}}{{{k}{D^{{\alpha _S}}}}}} \right)} \right)\overline {\cal H}\left( {x - {k}{D^{{\alpha _S}}}} \right)} \right)
\end{split}
\end{equation}
\normalsize \hrulefill \vspace*{-0.50cm}
\end{figure*}
\begin{figure*}[!t]
\setcounter{equation}{11}
\begin{equation}
\label{eq_MDR_4}
\begin{split}
 \widetilde R_i  &= \frac{1}{{\ln \left( 2 \right)}}\int\limits_0^{ + \infty } {\widetilde J_i \left( x \right)\exp \left( { - \Lambda _1 \left( {\left[ {0,x} \right)} \right) - \Lambda _2 \left( {\left[ {0,x} \right)} \right)} \right)\widehat \Lambda _{i,{\rm{LOS}}} \left( {\left[ {0,x} \right)} \right)dx}  \\
  &+ \frac{1}{{\ln \left( 2 \right)}}\int\limits_0^{ + \infty } {\widetilde J_i \left( x \right)\exp \left( { - \Lambda _1 \left( {\left[ {0,x} \right)} \right) - \Lambda _2 \left( {\left[ {0,x} \right)} \right)} \right)\widehat \Lambda _{i,{\rm{NLOS}}} \left( {\left[ {0,x} \right)} \right)dx}
\end{split}
\end{equation}
\normalsize \hrulefill \vspace*{-0.20cm}
\end{figure*}
\section{Rate Analysis} \label{ASE_RATE_Analysis}
In this section, we study the average rate of cellular networks by considering two case studies: 1) neither the spectrum nor the BSs (infrastructure) are shared among the operators and 2) both the spectrum and the BSs are shared among the operators. For ease of illustration, we focus our attention on a cellular network with two operators. The BSs of the operators are distributed according to two PPPs, which are denoted by $\Psi_i$ for $i=1,2$. The transmission bandwidth, deployment density and transmit power of the BSs of the $i$th operator are denoted by $W_i$, $\lambda_i$ and $P_i$, respectively. In what follows, in particular, the same notation as in the previous section is used and the subscript $i=1,2$ is used to distinguish between the two operators.

If neither spectrum nor infrastructure sharing are permitted between the two operators, the MTs are allowed to connect only to the BSs of their own operator. In this case, the transmission bandwidth and the transmit power are $W_i$ and $P_i$, respectively, where $i=1$ or $i=2$ depending on the operator being considered. If, on the other hand, both spectrum and infrastructure sharing are permitted, the MTs are allowed to connect to any available BS of both operators that provides the smallest path-loss. In this case, the transmission bandwidth is $W_1+W_2$ and the transmit power is $P_i$ where $i=1$ or $i=2$ depending on the serving operator. In the non-sharing case, the interference suffered by the MTs is lower but the transmission bandwidth is smaller. In the sharing case, on the other hand, the interference suffered by the MTs is higher but the transmission bandwidth is larger.

Based on these assumptions, let us consider the typical MTs that are subscribers of the first and of the second operator. The aggregate rate (in bits/sec) of both MTs can be formulated as $R_{\rm{nsh}} = W_1 \overline R_1 + W_2 \overline R_2$ and $R_{\rm{sh}} = 2(W_1+W_2) (\widetilde R_1 + \widetilde R_2)$ for the non-sharing and sharing setups, respectively, where $\overline R_i$ and $\widetilde R_i$ are the rates of the typical MT that is served by the $i$th operator for the non-sharing and sharing setups, respectively. Since the aggregate other-cell interference is different in the non-sharing and sharing case, in general, $\overline R_i \ne \widetilde R_i$ holds. $\overline R_i$ and $\widetilde R_i$ are computed in what follows.
\subsection{Computation of $\overline R_i$}
In order to compute $\overline R_i$, we use the MGF-based approach introduced in \cite{MDR_TCOMrate} and take the impact of LOS and NLOS links into account by using the same mathematical steps as in \cite{MDR__mmWave}. In particular, $\overline R_i$ can be formulated as follows:
\setcounter{equation}{7}
\begin{equation}
\label{eq_MDR_1}
\begin{split}
 \overline R_i  &= \frac{{1 }}{{\ln \left( 2 \right)}}\int\limits_0^{ + \infty } {\overline J_i\left( x \right)\exp \left( { - \Lambda_i \left( {\left[ {0,x} \right)} \right)} \right)\widehat \Lambda _{i,{\rm{LOS}}} \left( {\left[ {0,x} \right)} \right) dx}  \\
  &+ \frac{{1 }}{{\ln \left( 2 \right)}}\int\limits_0^{ + \infty } {\overline J_i\left( x \right)\exp \left( { - \Lambda_i \left( {\left[ {0,x} \right)} \right)} \right)\widehat \Lambda _{i,{\rm{NLOS}}} \left( {\left[ {0,x} \right)} \right) dx}
\end{split}
\end{equation}
where ${\Lambda_i \left( {\left[ {\cdot,\cdot} \right)} \right)}$, ${\Lambda _{i,{\rm{LOS}}} \left( {\left[ {\cdot,\cdot} \right)} \right)}$ and ${\Lambda _{i,{\rm{NLOS}}} \left( {\left[ {\cdot,\cdot} \right)} \right)}$ are the same as in \eqref{eq_inten1} and \eqref{eq_inten2} obtained by replacing $\lambda$ with $\lambda_i$, and $\overline J_i\left( \cdot \right)$ is defined as follows:
\setcounter{equation}{8}
\begin{equation}
\label{eq_MDR_2}
\overline J_i\left( x \right) = \int\limits_0^{ + \infty } {\frac{{P_i \exp \left( { - z} \right)}}{{P_i z + \sigma _{N,i}^2 x}}{\rm{\overline {MGF}}}_{I,i} \left( {\frac{{P_i }}{{\sigma _{N,i}^2 }}z;x} \right)dz}
\end{equation}
where $\sigma _{N,i}^2$ is the noise variance and:
\setcounter{equation}{9}
\begin{equation}
\label{eq_MDR_3}
{\rm{\overline {MGF}}}_{I,i} \left( {z;x} \right) = {\rm{\overline {MGF}}}_{I,i,{\rm{LOS}}} \left( {z;x} \right){\rm{\overline {MGF}}}_{I,i,{\rm{NLOS}}} \left( {z;x} \right)
\end{equation}
is the MGF of the aggregate other-cell interference, where ${\rm{\overline {MGF}}}_{I,i,{\rm{LOS}}} \left( {\cdot;\cdot} \right)$ and ${\rm{\overline {MGF}}}_{I,i,{\rm{NLOS}}} \left( {\cdot;\cdot} \right)$ are defined in \eqref{eq_MGF} shown at the top of this page.
\subsection{Computation of $\widetilde R_i$}
The computation of $\widetilde R_i$ follows a similar approach. In this case, however, we take into account that each MT is subject to the interference of both operators. More specifically, $\widetilde R_i$ can be formulated as provided in \eqref{eq_MDR_4} shown at the top of this page, where ${\Lambda_i \left( {\left[ {\cdot,\cdot} \right)} \right)}$, ${\Lambda _{i,{\rm{LOS}}} \left( {\left[ {\cdot,\cdot} \right)} \right)}$ and ${\Lambda _{i,{\rm{NLOS}}} \left( {\left[ {\cdot,\cdot} \right)} \right)}$ are the same as in \eqref{eq_inten1} and \eqref{eq_inten2} obtained by replacing $\lambda$ with $\lambda_i$, and $\widetilde J_i\left( \cdot \right)$ is defined as follows:
\setcounter{equation}{12}
\begin{equation}
\label{eq_MDR_5}
\widetilde J_i\left( x \right) = \int\limits_0^{ + \infty } {\frac{{P_i \exp \left( { - z} \right)}}{{P_i z + \sigma _{N,i}^2 x}}{\rm{\widetilde {MGF}}}_I \left(\frac{z}{\sigma _{N,i}^2};x \right)dz}
\end{equation}
as well as:
\setcounter{equation}{9}
\begin{equation}
\label{eq_MDR_6}
\begin{split}
 \widetilde{{\rm{MGF}}}_I \left( {z;x} \right) & = \widetilde{{\rm{MGF}}}_{I,1,{\rm{LOS}}} \left( {P_1 z;x} \right)\widetilde{{\rm{MGF}}}_{I,1,{\rm{NLOS}}} \left( {P_1 z;x} \right) \\
 & \times \widetilde{{\rm{MGF}}}_{I,2,{\rm{LOS}}} \left( {P_2 z;x} \right)\widetilde{{\rm{MGF}}}_{I,2,{\rm{NLOS}}} \left( {P_2 z;x} \right) \\
 \end{split}
\end{equation}
is the MGF of the aggregate other-cell interference, where $\widetilde {{\rm{MGF}}} _{I,i,S} \left( {z;x} \right) =  \overline{{\rm{MGF}}}_{I,i,S} \left( {z;x} \right)$ for $i=1,2$ and $S \in \{{\rm{LOS}}, {\rm{NLOS}}\}$.
\begin{table*}[!t]
\caption{Aggregate average rate (Mbits/sec).}
\centering
\begin{tabular}{|c|c|c|c|c|c|c|}
\hline
  &  $W_2/W_1=0.2$ &  $W_2/W_1=1$ & $W_2/W_1=2$ & $W_2/W_1=3$ & $W_2/W_1=4$ & $W_2/W_1=5$ \\
\hline
\hline
  & \multicolumn{6}{c|}{Non-Sharing System Setup} \\
\hline
$P_2/P_1=0.2$ & 80.7 & 134.3 & 203.1 & 270.0 & 335.9 & 399.2\\
$P_2/P_1=1.0$ & 80.7 & 134.5 & 201.7 & 268.8 & 335.7 & 402.7\\
$P_2/P_1=5.0$ & 80.7 & 134.5 & 203.5 & 269.0 & 336.2 & 403.4\\
\hline
\hline
  & \multicolumn{6}{c|}{Spectrum and Infrastructure Sharing System Setup} \\
\hline
$P_2/P_1=0.2$ & 156.1 & 259.9 & 389.4 & 518.5 & 646.9 & 774.7\\
$P_2/P_1=1.0$ & 161.4 & 268.8 & 403.0 & 536.9 & 670.9 & 804.5\\
$P_2/P_1=5.0$ & 156.2 & 260.3 & 390.3 & 520.3 & 650.4 & 780.3\\
\hline
\end{tabular}
\label{table1}
\end{table*}
\section{Numerical Results} \label{Results}
In this section, we show numerical results that are aimed to compare the aggregate average rate of cellular networks with and without spectrum and infrastructure sharing. The following setup is assumed: $k=\left({4\pi f_c}/{c}\right)^2$, where $f_c=2.1$ GHz is the carrier frequency and $c$ is the speed of light, $\sigma^2_{N,i}[{\rm{dB}}]=-174+10\log_{10}(W_i)+N_f$ dB, where $N_f=10$ dB is the noise figure, $D=109.8517$ m, $\alpha_{\rm{LOS}}=2.5$, $\alpha_{\rm{NLOS}}=3.5$, $q_{\rm{LOS}}^{[0,D)}=0.7195$ and $q_{\rm{LOS}}^{[D,\infty)}=0.0002$ \cite{lu2015stochastic}.

Due to the presence of LOS and NLOS links, an optimal value for the density of the BSs that maximizes the average rate of both non-sharing and sharing setups exists \cite{lu2015stochastic}. Based on the proposed mathematical frameworks, these optimal values of $\lambda$ are computed for different values of the transmission bandwidths and transmit powers of both operators. Then, the corresponding optimal aggregate average rates are obtained. The results of this study are summarized in Table \ref{table1}, which highlights the potential advantages of spectrum and infrastructure sharing in cellular networks.
\section{Conclusion} \label{Conclusion}
In this paper, we have studied multi-operator cellular networks that are allowed to share spectrum licenses and infrastructure elements. We have proposed a mathematical framework for system-level analysis and optimization, which exploits tools from stochastic geometry for modeling the locations of cellular base stations. Our numerical results show that spectrum and infrastructure sharing are capable of increasing the aggregate rate of multi-operator cellular networks.
\section*{Acknowledgment}
This work is supported in part by the European Commission through the H2020-ETN-5Gwireless project under grant 641985 and through the H2020-ETN-5Gaura project under grant 675806.

\begin{thebibliography}{99}
%
\bibitem{andrews2014will} J. G. Andrews, S. Buzzi, W. Choi, S. Hanly, A. Lozano, A. Soong, and C. Zhang,
            ``What will 5G be?'',
             \emph{IEEE J. Sel. Areas in Commun.},
             vol. 32, no. 6, pp. 1065-1082, June 2014.
%
\bibitem{DaSilva} J. Kibilda, P. Di Francesco, F. Malandrino, and L. A. DaSilva,
            ``Infrastructure and spectrum sharing trade-offs in mobile networks'',
             \emph{IEEE Dynamic Spectrum Access Networks},
             pp. 1-10, Oct. 2014.
%
\bibitem{jorswieck2014spectrum} E. Jorswieck, L. Badia, T. Fahldieck, E. Karipidis, and J. Luo,
             ``Spectrum sharing improves the network efficiency for cellular operators'',
             \emph{IEEE Commun. Mag.},
             vol. 52, no. 3, pp. 129-136, Mar. 2014.
%
\bibitem{janssen2014expiration} T. Janssen, R. Litjens, and K. Sowerby,
             ``On the expiration date of spectrum sharing in mobile cellular networks'',
             \emph{Int. Symposium on Modeling and Optimization in Mobile, Ad Hoc, and Wireless Networks},
             pp. 490-496, 2014.
%
\bibitem{peha2009sharing} J. Peha,
             ``Sharing spectrum through spectrum policy reform and cognitive radio'',
             \emph{Proc. of the IEEE},
             vol. 97, no. 4, pp. 708-719, Apr. 2009.
%
\bibitem{MDR_TCOMrate} M. Di Renzo, A. Guidotti, and G. E. Corazza,
              ``Average rate of downlink heterogeneous cellular networks over generalized fading channels -- A stochastic geometry approach'',
             \emph{IEEE Trans. Commun.},
             vol. 61, no. 7, pp. 3050--3071, July 2013.
%
\bibitem{lu2015stochastic} W. Lu and M. Di Renzo,
             ``Stochastic geometry modeling of cellular networks: Analysis, simulation and experimental validation'',
             \emph{Int. Conference on Modeling, Analysis and Simulation of Wireless and Mobile Systems},
             pp. 179-188, Nov. 2015.
%
\bibitem{MDR_COMML2014} M. Di Renzo and W. Lu,
             ``The equivalent-in-distribution (EiD)-based approach: On the analysis of cellular networks using stochastic geometry'',
             \emph{IEEE Commun. Lett.},
             vol. 18, no. 5, pp. 761-764, May 2014.
%
\bibitem{Peng_TCOM} M. Di Renzo and P. Guan,
              ``A mathematical framework to the computation of the error probability of downlink MIMO cellular networks by using stochastic geometry'',
             \emph{IEEE Trans. Commun.},
             vol. 62, no. 8, pp. 2860--2879, Aug. 2014.
%
\bibitem{MDR_COMMLPeng} M. Di Renzo and P. Guan,
             ``Stochastic geometry modeling of coverage and rate of cellular networks using the Gil-Pelaez inversion theorem'',
             \emph{IEEE Commun. Lett.},
             vol. 18, no. 9, pp. 1575-1578, Sep. 2014.
%
\bibitem{Wei_TCOM} M. Di Renzo and W. Lu,
              ``Stochastic geometry modeling and performance evaluation of MIMO cellular networks using the equivalent-in-distribution (EiD)-based approach'',
             \emph{IEEE Trans. Commun.},
             vol. 63, no. 3, pp. 977-996, Mar. 2015.
%
\bibitem{WeiRelays_TCOM} W. Lu and M. Di Renzo,
              ``Stochastic geometry modeling and system-level analysis \& optimization of relay-aided downlink cellular networks'',
             \emph{IEEE Trans. Commun.},
             vol. 63, no. 11, pp. 4063-4085, Nov. 2015.
%
\bibitem{MDR__mmWave} M. Di Renzo,
             ``Stochastic geometry modeling and analysis of multi-tier millimeter wave cellular networks'',
             \emph{IEEE Trans. Wireless Commun.},
             vol. 14, no. 9, pp. 5038-5057, Sep. 2015.
%
\bibitem{WLu__mmWave} W. Lu and M. Di Renzo,
             ``Stochastic geometry modeling of mmWave cellular networks: Analysis and experimental validation'',
             \emph{IEEE Int. Workshop on Measurement and Networking},
             pp. 1-5, Oct. 2015.
%
\end{thebibliography}
\end{document}